\documentclass{vgtc}                          

\ifpdf
  \pdfoutput=1\relax                   
  \usepackage{graphicx}                
  \DeclareGraphicsExtensions{.pdf,.png,.jpg,.jpeg} 
\else
  \usepackage{graphicx}                
  \DeclareGraphicsExtensions{.eps}     
\fi%

\graphicspath{{figures/}{pictures/}{images/}{./}} 

\usepackage{microtype}                 
\PassOptionsToPackage{warn}{textcomp}  
\usepackage{textcomp}                  
\usepackage{mathptmx}                  
\usepackage{times}                     
\usepackage{cite}                      
\usepackage{tabu}                      
\usepackage{booktabs}                  

\usepackage[T1]{fontenc}
\usepackage[utf8]{inputenc}
\usepackage{amsfonts}

\onlineid{0}

\vgtccategory{Research}

\vgtcinsertpkg

\title{Optimization and Manipulation of Contextual Mutual Spaces for Multi-User Virtual and Augmented Reality Interaction}

\author{Mohammad Keshavarzi\thanks{e-mail: mkeshavarzi@berkeley.edu} ${\;  }^{1, 2}$ 
\and Allen Y. Yang\thanks{e-mail: yang@eecs.berkeley.edu} ${\;  }^{ 1}$ %
\and Woojin Ko\thanks{e-mail: woojin\_ko@berkeley.edu}  ${\;  }^{ 1}$
\and Luisa Caldas\thanks{e-mail: lcaldas@berkeley.edu} ${ \; }^{ 2}$}

\affiliation{\scriptsize  ${}^{1}$ FHL Vive Center for Enhanced Reality, University of California, Berkeley \\ 
${}^{2}$ XR Lab, Department of Architecture, University of California, Berkeley}

\abstract{Spatial computing experiences are physically constrained by the geometry and semantics of the local user environment. This limitation is elevated in remote multi-user interaction scenarios, where finding a common virtual ground physically accessible for all participants becomes challenging. Locating a common accessible virtual ground is difficult for the users themselves, particularly if they are not aware of the spatial properties of other participants. In this paper, we introduce a framework to generate an optimal mutual virtual space for a multi-user interaction setting where remote users' room spaces can have different layout and sizes. The framework further recommends movement of surrounding furniture objects that expand the size of the mutual space with minimal physical effort. Finally, we demonstrate the performance of our solution on real-world datasets and also a real HoloLens application. Results show the proposed algorithm can effectively discover optimal shareable space for multi-user virtual interaction and hence facilitate remote spatial computing communication in various collaborative workflows.
}

\CCScatlist{
\CCScatTwelve{Computing methodologies}{Computer graphics}{Graphics systems and interfaces}{Mixed / augmented reality};
\CCScatTwelve{Human-centered computing}{Human computer interaction }{Interaction paradigms}{Collaborative interaction};
\CCScatTwelve{Applied computing}{Decision analysis}{Multi-criterion optimization and decision-making}{};
\CCScatTwelve{Theory of computation}{Mathematical optimization}{Optimization with randomized search heuristics}{Evolutionary algorithms}
}

\begin{document}

\firstsection{Introduction}
\maketitle

The emerging fields of augmented reality (AR) and virtual reality (VR) have introduced a large number of exciting applications in tele-communication, immersive collaboration, and social media where multiple users can share a virtual environment. While much work has been done on 3D capturing methods, real-life avatar modeling, and virtual social platforms, one key challenge in AR/VR immersion is the scene understanding of the users' surrounding spaces and the question of how to optimally utilize them for immersive tasks.

More specifically, acquiring an accessible 3D workspace is a prerequisite for a virtual or augmented immersion experience. Furthermore, the augmentation of the virtual data in the physical space must be compatible with the contextual properties of the physical space, such as a floor that is standable, a chair that is sittable, and a wall that is also a physical barrier of virtual interactions. For many 6 degrees-of-freedom (DOF) VR applications, the user will often be asked to manually initiate a block of free space where the VR immersion can be assumed to be safe. Inferencing the above contextual information for both AR and VR can be readily done using several well-established 3D modeling algorithms in computer vision. Current AR devices, such as the HoloLens or MagicLeap, integrate such algorithms to estimate the layout of the space, including floors, walls, and ceilings, and typical furniture objects such as tables and chairs. In this paper, we assume such contextual information of individual spaces to be available via either a manual or algorithmic process.

However, in scenarios where an immersive experience involves multiple users, understanding of spatial constraints is elevated to that of all involved users. Since different users may participate in the immersive experience from their own spaces, which can hold very contrasting contextual properties, a consensus must be established to identify a mutual space that respects the spatial constraints of all the participants. Yet, having users manually identify such a mutual space would be imprecise and labor intensive, especially when considering the fact that it would be difficult for a user to be aware of the contextual properties of the other users' spaces. Without more effective and efficient solutions, the establishment of a contextual mutual space will be a bottleneck for multi-user immersion experiences.

Motivated by this challenge, we present in this paper a novel method to optimize contextual mutual spaces in a multi-user immersion setting. Our method relies on existing semantic scene maps to identify shareable functional spaces, and is general enough to optimize contextual mutual spaces even when the users' spaces have very different layouts and sizes (see results in Figure \ref{fig:HoloLens}). For illustration purposes, we will use standable and sittable as the two exemplary contextual functions to develop our method, and the proposed solution is compatible with other contextual functions that can be modeled by the same mathematical framework. The method formulates an optimization problem to seek the maximal mutual spaces. Furthermore, if one can assume the users have the freedom to rearrange furniture objects on the floor, we introduce a more delicate optimization process to further increase the mutual space's size while balancing the users' efforts to physically move the objects as another constraint. To effectively solve the above two problems, we propose to use a genetic algorithmic approach. Clearly, we believe other comparable algorithms that optimize these NP-Hard problems are equally effective. Nevertheless, our results validate a new approach capable of automatically recommending contextual mutual space to multiple participants of virtual immersion experiences in AR/VR applications.

We believe our proposed framework can play a role in facilitating remote workplace practices and virtual collaborations by decreasing the spatial requirements for tele-presence systems. Instead of setting up large open spaces required for such workflows, our system would allow users to join from their personal spaces, with minimum modifications to their surrounding environment. Physical and virtual re-arrangements would be optimized based on the number of participant and their local environments. In AR experiences, the topological relationship and line of sight between all participants would be maintained without any conflicts between remote users and local physical obstacles, while in VR, our system can recommend spatial modifications and provide the required interaction area between multiple users.

\section{Related Work}

Immersive AR/VR systems have been widely explored for remote tele-presence applications, providing real-time capture, transmission and display between participants of the platform \cite{Fuchs2014,Beck2013,Kuster2012}. Using an array of cameras \cite{towles20023d,kurillo2008immersive,tanikawa2005real} or depth sensors monitoring the capture space \cite{Pejsa:2016:REL:2818048.2819965,jones2014roomalive,matusik20043d,maimone2012real}, holographic replicas or avatars \cite{wei2019vr, Lombardi} of the virtual participants are projected in pre-defined local spaces. Such projections have been extensively developed using situated autostereo \cite{nagano2013autostereoscopic,matusik20043d}, volumetric \cite{jones2009achieving}, lightfield \cite{balogh2010real}, cylindrical \cite{kim2012telehuman}, and holographic \cite{blanche2010holographic} displays. However, participants of such systems are mainly stationed in predefined spaces \cite{zhang2013viewport,benko2012miragetable,maimone2011encumbrance,wen2000toward} to avoid any geometrical conflicts with surrounding features in the projected space. Such approach limits free-form motion of the participants within each other's location, an important factor for achieving co-located presence.

\begin{figure}
  \includegraphics[width=0.95\columnwidth]{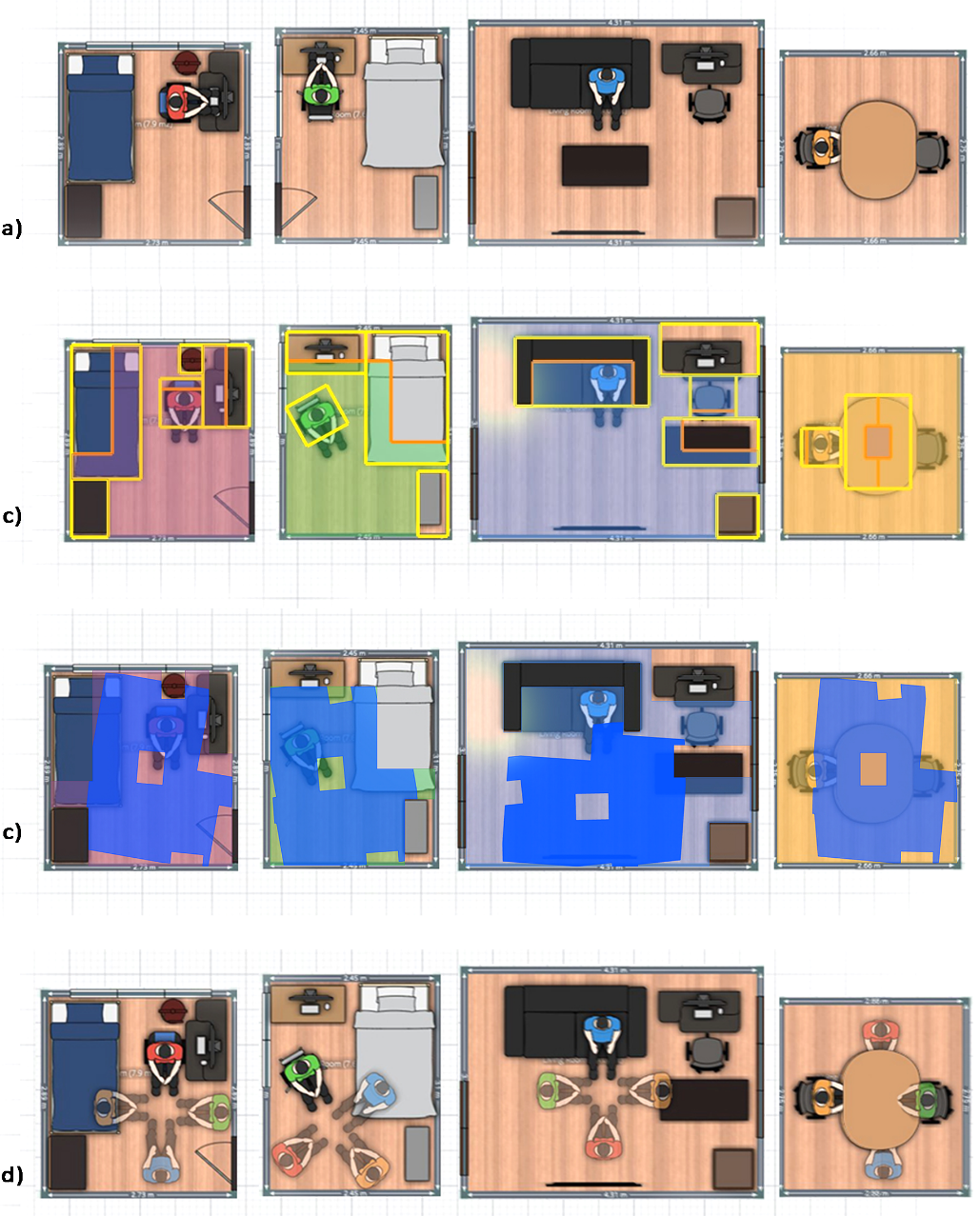}
  \caption{Abstract illustration of our proposed framework a) initial settings with different spatial restrictions b) semantic segmentation defining standable (yellow boundaries) and sittable (orange boundaries) areas c) search for mutual sittable space (this step can be before, after or simultaneous with object repositioning) d) virtual arrangement of avatars with deterministic line of sight of all participants.}~\label{fig:conceptFig}
\end{figure}

The importance of free-form user movement and the ability to preserve mobility-based communication features such as walking, gestures, and head movement have been studied greatly in the context of co-located collaboration \cite{bardram2005activity,keshavarzi2019affordance,luff1998mobility, caldas2019design}. Another vital aspect of sharing mutual space is described in Clark's work as grounding \cite{clark1991grounding}. Grounding in communication (or common ground) is a concept that comprises the collection of "mutual knowledge, mutual beliefs, and mutual assumptions" that is essential for communication between two people. Successful grounding in communication requires parties to coordinate both the content and process \cite{LEE200121}. As content in spatial computing can also involve the surrounding space itself, providing a common virtual ground can be critical to allow all communication features to be reflected correctly.

More recent examples have explored how tele-presence can be conducted with less spatial constraints, allowing fluid user motion in both ends of the communication. Works of \cite{gross2003blue} and \cite{beck2013immersive} are examples of such systems where users and their local interaction spaces are continuously captured using a cluster of registered depth and color cameras. However, these systems use stereoscopic projection which limits the ability for remote and local users to  access each others space. Instead, spaces are virtually disconnected and interaction occurs through a window from one space into the other. Meanwhile, the Holoportation system introduced by Orts-Escolano et al. allows bilateral tele-presence between participants where participants share a common virtual ground \cite{Orts-Escolano2017}. Their system allows the remote user to be rendered into the local user's space as an avatar while the local user appears as an avatar in the remote user's space as well. Such an approach is also seen in \cite{maimone2013general}, where the remote and local users do not share the same functional layout of rooms, but they are calibrated in order to provide the required mutual virtual ground between users.

While tele-presence systems via shared spaces present novel workflows for capturing and projecting virtual avatars, the issue of avoiding physical and virtual conflicts within the shared spaces is still an open challenge. Narang et al. \cite{narang2018simulating} developed a system which generates non-colliding movements for human-like agents interacting with other agents or avatars in a virtual environment. Lang et al. \cite{lang2019virtual} integrate scene semantics with a Markov chain Monte Carlo optimization method to find optimized locations for placing virtual agents close to a single user. Such approach addresses the spatial limitations of a single user, but not multiple constraints generated by multiple remote user. The work of Lehment et al. \cite{Lehment2014} may be the closest work to this paper, which proposes an automated method to align remote environments so that they minimize discrepancies in room obstacles and physical barriers. However, the method is limited to two spaces and uses a brute force search to calculate the consensus space between participants. Our method formulates rigorous optimization problems to search and manipulate a potentially unlimited number of spaces in order to find a mutual spatial boundary.

For virtual reality environments, techniques in redirected walking \cite{Razzaque2001} also aim to resolve the possible conflicts of virtual and physical surroundings. While the focus is mainly on providing a natural locomotion of a local user, such techniques use subtle (redirected without the user’s knowledge) \cite{bruder2011tuning, bolte2015subliminal} or overt (detectable by the user) \cite{interrante2007seven, williams2007exploring, peck2011evaluation} strategies to manipulate the mapping between the user’s real and virtual translation and rotation, resulting the user to avoid interference with edges of the usable space or physical obstacles. Architectural manipulation of virtual spaces has also been investigated by re-arranging virtual elements in blind-spots \cite{Suma2011} or implementing self-overlapping \cite{Suma2012} and flexible virtual spaces \cite{Vasylevska2013}. However, redirected walking techniques may introduce simulator sickness \cite{Nilsson2018}, interfere with spatial  memory \cite{williams2007exploring}, and lead to higher cognitive load than real world locomotion  \cite{bruder2015cognitive}. Furthermore, while such strategies can be applied in VR environments, they cannot generally apply for AR experiences due to the see-through nature of AR. Even so, the ability to efficiently manipulate the real-world surroundings introduced by our system would provide more spatial freedom, especially in remote mulit-user scenarios, before applying multi-user redirecting walking techniques. 

Part of our proposed system intends to determine an optimal arrangement of discrete spatial elements within a room. Such practice is often referred to as floorplanning \cite{eisenmann1998generic}. Automated floorplanning methodologies have been widely investigated in architectural space layouts, construction \cite{Cheng_1992, Tommelein_Levitt_Hayes-Roth_Confrey_1991, Osman_Georgy_Ibrahim_2003}, electronic design \cite{nakatake1997module, Chang_Chang_Wu_Wu_2000,gwee1999ga}, and industrial operation research \cite{Abdinnour-Helm_Hadley_2000}. Floorplanning aims to achieve a defined functional goal by efficiently generating and evaluating possible spatial combinations while addressing the geometrical and topological constraints of the spatial elements \cite{Jo_Gero_1998}. In electronic physical design floorplanning, proposed methodologies mostly aim at optimizing chip area and wirelengths to reduce interconnections and improve timing \cite{kahng2011vlsi}. In construction site layout and planning, optimizing the interaction between facilities, such as total inter-facility transportation costs and frequency of inter-facility trips can also be implemented as objective functions \cite{Osman_Georgy_Ibrahim_2003}. In our proposed framework, we similarly integrate an objective function whose goal is to minimize the amount of effort required to move surrounding furniture while maximizing the area of the mutual virtual ground among all participants. 

In floorplanning, various representation methods of spatial arrangements are coupled with optimization engines to efficiently search through all possible combinations of spatial elements. Floorplanning representations are generally divided into two main categories: slicing and non-slicing representations \cite{Wang:2009:EDA:2843514}. In slicing methodologies, the floor plan is recursively bisected until each part consists of a single module \cite{Wong:1986:NAF:318013.318030}. Non-slicing representation are utilized for more general use cases where no recursive bisection of a certain area takes place \cite{Guo_Cheng_Yoshimura_2003, ma2001vlsi,lin2005tcg}.
Multiple studies have integrated these representations with various optimization algorithms such as Simulated Annealing (SA) \cite{kirkpatrick1983optimization, kiyota2005simulated, Wong:1986:NAF:318013.318030}, Genetic Algorithms (GA) \cite{rebaudengo1996gallo,nakaya2000adaptive, lin2002efficient,gwee1999ga,xiaogang2002vlsi} and Particle Swarm Optimization (PSO) \cite{sun2006floorplanning,chen2008vlsi,kaur2014enhanced,sowmya2013minimization,moni2009vlsi}. More recently, by applying learning based algorithms, hybrid neural networks[7] and annealed neural networks  have been used to identify optimal site layout and solve construction site-level problems.[8]

\section{Methodology}
Our solution consists of the following four steps:
(i) Semantic segmentation of surrounding environments;
(ii) Topological scene graph generation;
(iii) Mutual space identification;
(iv)  Optionally, manipulation of ground objects to further maximize the mutual space. In this section, we will elaborate on the details of the four steps. To start, we will define the terminologies and notations used in the paper.

Given a closed 3D room space in $\mathbb{R}^3$, one can project its enclosure, i.e., floors, ceilings, and walls, via an orthographic projection to form a 2D projection, which is commonly known as the floor plan of the space. If we assign the $(x, y)$ coordinates on the floor-plan plane and the $z$ coordinate perpendicular to the floor-plan plane, simplifying our optimization problems on to the $(x, y)$ plane significantly reduces the complexity of our algorithms. It also implies an assumption that there is no overlap between two objects on the $(x, y)$ plane but with different $z$ values. Nevertheless, we believe such simplification is reasonable for analyzing the majority of room structures and thus does not compromise the generality of our analysis provided herein.

Hence, we define for each user $i$ their own room space expressed as a 2D floor plan as $R_i$. Each $k$-th object (e.g., furniture) in $R_i$ is denoted as $O_{i,k}$.The collection of all $n_i$ objects in $R_i$ is denoted as $\mathcal{O}_i= \{O_{i,1}, O_{i,2},... O_{i,n_i}\}$.  $\bar{O}_{i,k}$ represents the boundary of the object $O_{i,k}$. Similarly, $\bar{R}_i$ represents the boundary of the room $R_i$. Finally, we define the area function as $K(O)$.

\begin{figure}
\centering
   \includegraphics[width=0.9\columnwidth]{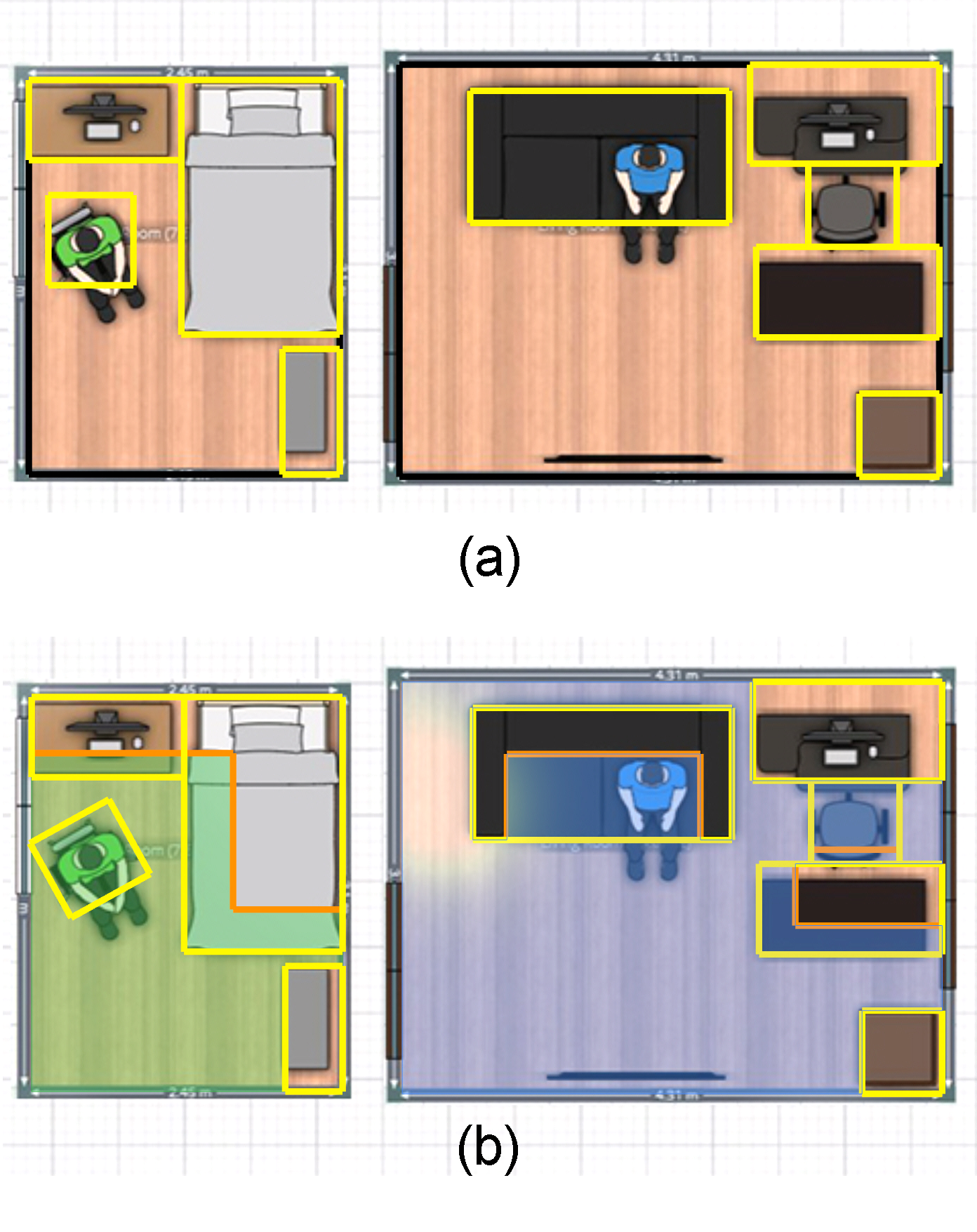}
  \caption{Comparison between available (a) standing only and (b) standing and sitting area in rooms. }~\label{fig:standSitting}
\end{figure}

\subsection{Semantic Segmentation}
Given the measurement of the surrounding physical environments as large sets of point cloud data, one can take advantage of the semantic segmentation methods widely investigated in computer vision literature \cite{Qi2016,Liu2018,Armeni2016} to segment their spatial boundaries and obtain their geometric properties, such as dimensions, position and orientation, object classification, functional shapes, and their weights. In doing so, we can convert the 3D point cloud data to labeled objects $O_{i,k}$ with a bounding box as $\bar{O}_{i,k}$.


Additionally, in this paper we exclude lightweight objects (such as pillows, alarm clocks, laptops, etc.) positioned on larger furniture. This is to simplify our calculations in the next steps as we assume these lightweight objects can be easily moved by the users and do not need to be considered in the optimization criteria. Such classification is dependent on the output labeled object categories above. 

In the experiment section below, since the implementation of a computer vision algorithm for semantic segmentation is not the main focus of this paper, we will directly integrate a modified version of Matterport 3D \cite{Chang2018} object classifier in our system. This module can be replaced with any other robust semantic segmentation algorithms, as long as they provide bounding box coordinates for each object category. In a companion Matterport 3D \cite{Chang2018} dataset, out of 1,659 unique text labels, we classify 134 of the labels as lightweight objects and filter their corresponding bounding box from our workflow.

Figure \ref{fig:standSitting}(a) illustrates the result of semantic segmentation of two room spaces projected onto the $(x,y)$-plane.

\subsection{Topological Scene Graph}

After identifying the bounding box, orientation, and category type of each object in the scene $R_i$, a topological graph is readily generated that describes the relationship and constraints of the objects between one each other within $R_i$. This step will allow us to identify usable spatial functions such as standing in virtual immersion, located between the objects. We categorize this type of functions as \emph{standalone spatial functions}, and their spaces are called \emph{standalone spaces}.

A topological scene graph will also allow us to identify other spatial functions on the objects themselves such as sitting on a chair and working on a table. But note that such functions as sitting or working are also constrained by the distances between the object that performs the function and its adjacent other objects. For example, a side of the table can not be utilized for working purposes if that side is adjacent to other furniture or building elements (such as walls, doors, etc.). We categorize this type of functions as \emph{auxiliary spatial functions}, and their spaces are called \emph{auxiliary spaces}. 

In this paper, we will use two spatial functions \emph{standable} and \emph{sittable} as an example to demonstrate how to integrate both standalone spatial functions and auxiliary spatial functions in the optimization of contextual mutual spaces for mutli-user interaction in AR/VR. 

Finally, we emphasize that standalone spaces and auxiliary spaces are not mutually exclusive. For example, in this paper, we will classify that a standable space can be assumed to be sittable as well. However, the vice versa may not be true. For example, a portion of a sittable space involves a part of a bed object, which we will not assume to be standable. Such contextual constraints can be highly customizable based on the content of the AR/VR application. But the framework that we are introducing in this paper is general enough to accommodate other contextual interpretations of the standalone spatial functions and auxiliary spatial functions.


In our implementation, we use a doubly-linked data structure to construct the graph. For each side face of an object's bounding box we define the closest adjacent objects to the face and calculate the distance between the object and the specified face. This information would be stored at the object level, where topological distances and constraints are referenced using pointers. 

Mathematically, for each object $O_{i,k}$, we define the function $\delta X_{\max}(O_{i,k})$ as the shortest distance between the points in $O_{i,k}$ that have the maximal $x$ value and the other objects including $\bar{R}_i$. Similarly, we define the functions  $\delta X_{\min}(\cdot)$, $\delta Y_{\max}(\cdot)$, and $\delta Y_{\min}(\cdot)$.

\subsection{Mutual Space Identification}
In this step, we will identify the geometrical boundaries of available spaces in each room and then align the calculated boundaries of all rooms to achieve maximum consensus on mutual spaces.

First, using the geometrical and topological properties extracted in previous steps, we are ready to calculate available spaces in each room based on two categories, namely, the standalone spaces and auxiliary spaces. Specifically, we will formulate the calculation of the two most typical spatial functions as examples again, namely, standable and sittable.

\subsubsection{Standable Spaces}

Standing spaces consist of the volume of the room in which no object located within a human user's height range is present. In such spaces, user movement can be performed freely without any risk of colliding with an object in the surrounding physical environment. Activities such as intense gaming or performative arts can be safely executed within these boundaries. Such spaces are also suitable for virtual reality experiences, where users may not be aware of the physical surroundings.

We calculate the available standing space ($S$) for room $R_i$ simply as follows:

\begin{equation}
 S_{i} =  R_i - \bigcup_{k=1}^{n_i}O_{i,k}.
\end{equation}

\subsubsection{Sittable Spaces}
The calculation of maximal sittable spaces is more involved than that of the standable spaces above. As we mentioned before, sittable spaces normally extend the standable spaces by adding areas where humans are able to sit on. Furniture types such as sofas, chairs, and beds include sitting areas that can extend usable spaces of a room for social functions such as general meetings, design reviews, and conference calls. 

To start, we define a sittable threshold $\varepsilon(O_{i,k})$ to calculate the sittable area within the bounding box of the object $O_{i,k}$. In other words, 
$\varepsilon(O_{i,k})$  is the maximum distance inward from an edge of the object's bounding box that can be comfortably sit on. We use measurements from \cite{ramsey2007architectural} to define the $\varepsilon$ of each furniture type. If object $O$ is classified as non-sittable, then $\varepsilon(O) = 0$. 

Therefore, we can first calculate the non-sittable area of an object $O$ as
\begin{equation}
N(O)\doteq \{\forall p\in O: B(p,\varepsilon(O)) \cap  O =  B(p,\varepsilon(O))\},
\end{equation}
where $B(p, \varepsilon(O))$ is a sphere in $\mathbb{R}^2$ centered at $p$ and with radius $\varepsilon(O)$.

We note that sittable spaces do not necessarily comprise only objects to be sit on, but rather describe an area where a sittable object can be placed in. For example, while an individual may not be able to comfortably sit on the top of the table, but the foot space below the table can be considered as sittable space. Therefore, in such context the sittable area of the room is always larger than its standable area. 

Moreover, sittable areas of each object in the room is constrained by the topological positioning of the object. If any of the object's boundaries is adjacent to a non-sittable object (such as a wall, bookshelf, etc) or does not contain enough standable area between itself and a non-sittable object, the sittable area of the side of the face should be excluded. For instance, if a table is positioned in the center of a room, with no other non-sittable object around it, the sittable area would be calculated by applying the sittable threshold to all four sides of the table's boundaries. However, if the table is positioned in the corner of the room, then there will be no sittable area accumulated for the sides that are adjacent to the wall.   

\begin{figure}
\centering
  \includegraphics[width=0.95\columnwidth]{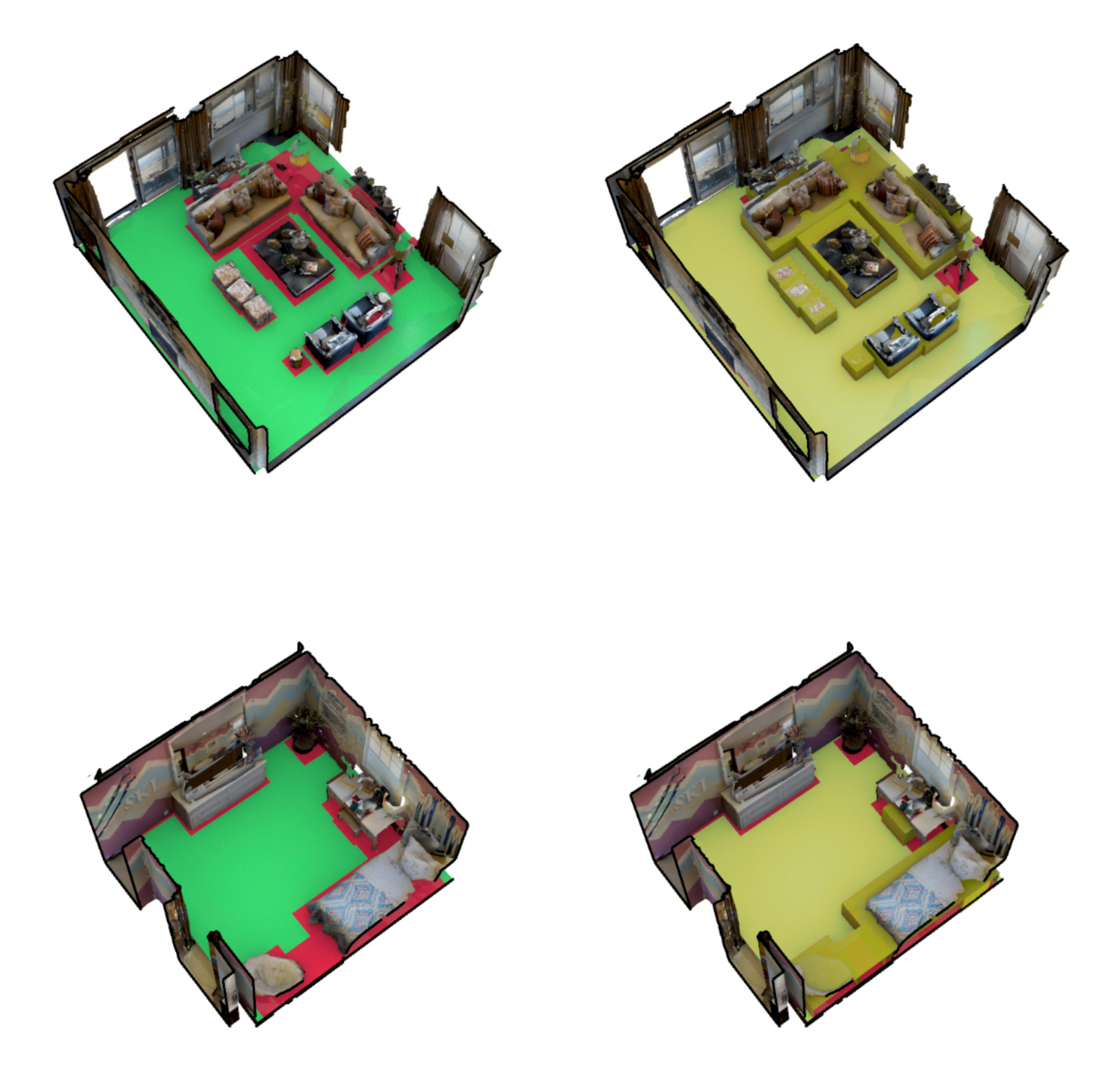}
  \caption{Standable (green), non-standable (red) and sittable spaces (yellow) for two example scenes from the Matterport 3D dataset.}~\label{fig:standSitMat}
\end{figure}

To simplify our calculation, we define a surrounding boundary threshold $\varrho(O)$ for object $O$, which measures the distance from any object's boundary point outward that allows that point to remain part of the sittable space of the object. In other words, if the boundary point is close to other objects or the room boundary within distance $\varrho$, then that point can not be sit on. $C(O_{i,k})$ defined below collects all such points for exclusion from $O_{i,k}$ in room $R_i$:
\begin{equation}
    \begin{tabular}{r l}
    $ C(O_{i,k}) =$ & 
    $ \{\forall p\in O_{i,k}: B(p,\varepsilon(O_{i,k})+\varrho(O_{i,k})) \cap \bar{R_i}\neq \emptyset $ \\ 
    & $ \mbox{or } B(p,\varepsilon(O_{i,k})+\varrho(O_{i,k})) \cap O_{i,h} \neq \emptyset, h\neq k \}$
    \end{tabular}
\end{equation}
where $\emptyset$ denotes the empty set. Therefore, the sittable space of each object $O$ is simply defined as
\begin{equation}
A(O) = O - N(O) \cup C(O).
\end{equation}

Finally, the total sittable space $A(R_i) $ for the room $R_i$ is
\begin{equation}
    A(R_i) = \bigcup_{k=1}^{n_i}A(O_{i,k}) + A(S_i).
\end{equation}


Figure \ref{fig:standSitMat} illustrates two example rooms and compares their standing and sitting areas.

\subsubsection{Maximizing Mutual Spaces}
Now we consider an immersive experience where there are $m$ subjects and therefore $m$ room spaces $(R_1, R_2, \cdots, R_m)$, respectively. Then, in the $(x,y)$-coordinates, we define a rigid-body motion in $\mathbb{R}^2$ as $G(F, \theta)$, where $\theta$ describes a translation and a rotation. 

If we want to maximize a mutual standable space, we can apply one $G(S_i, \theta_i)$ to each individual standable space $S_i$ for the $i$-th user. The optimal rigid body motion then maximizes the area of the interaction space:
\begin{equation}
    (\theta_1^*, \cdots, \theta_m^*)=\arg\max K(\bigcap_{i=1}^{m} G(S_i, \theta_i)).
\label{eq:optimal-rigid-body-motion}
\end{equation}



Then the maximal mutual standable space can be calculated as
\begin{equation}
    M_S(R_1,\cdots, R_m) = \bigcap_{i=1}^{m} G(S_i, \theta_i^*)
    \label{eq:max-standable-space}
\end{equation}

Similarly, one can calculate the maximal mutual sittable space $M_A(R_1,\cdots, R_m)$ by substituting the rigid body motions in (\ref{eq:max-standable-space}) that maximizes their intersection area function in (\ref{eq:optimal-rigid-body-motion}).

\begin{figure}
\centering
  \includegraphics[width=0.82\columnwidth]{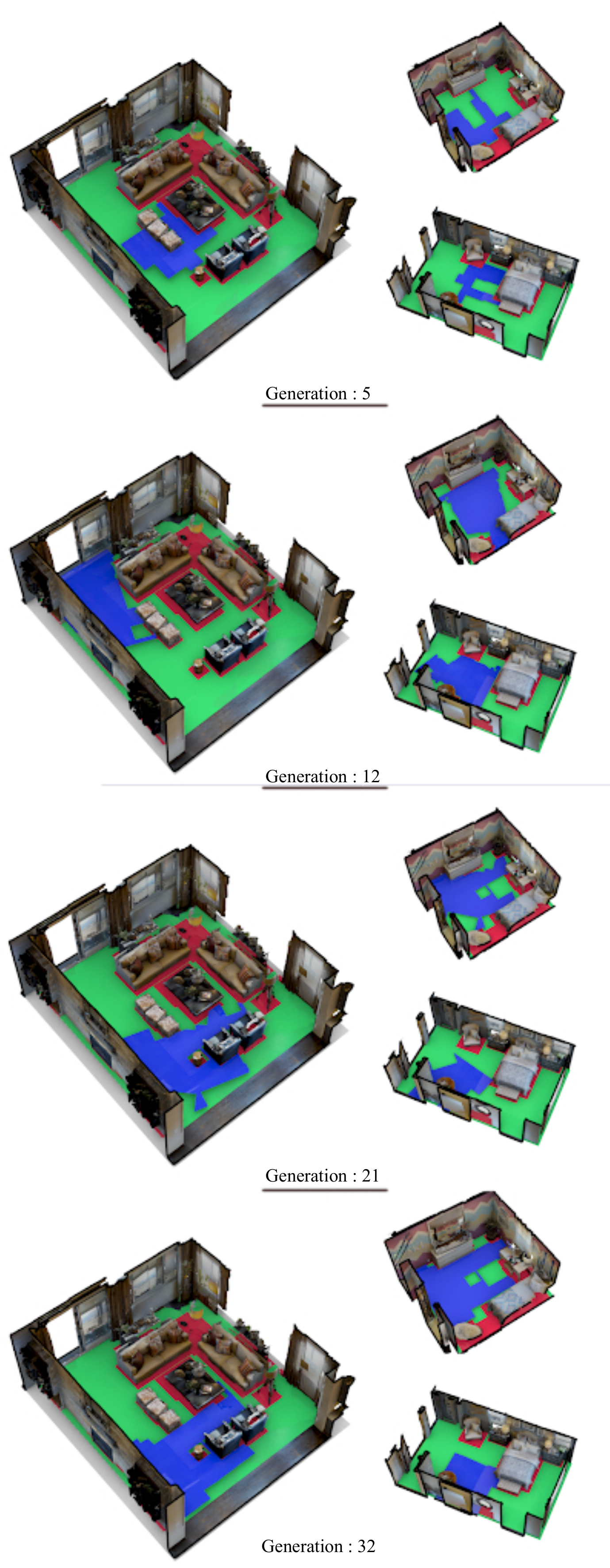}
  \caption{Mutual Spatial boundaries (blue) for different generations of the search mechanism. The green area indicates standable spaces and the red area indicates non-standable spaces. The result shows that the optimized mutual standable space increases over generations.}~\label{fig:MutSpace}
\end{figure}

\subsection{Furniture movement optimization}
In the event where individual spaces $R_i$ include movable furniture, additional optimization can be considered to potentially increase the maximal mutual spaces. Diverging from merely considering rigid-body motions to transform just the coordinate representation of the spaces, we consider moving furniture objects in space, which has an additional cost of human effort. Consequently, we will formulate this effort as part of our optimization objective.

More specifically, given a rigid-body motion $G$, we definite $\|G\|_t$ as the Euclidean distance of its translation vector. Then we define
\begin{equation}
    E = w\|G\|_t,
\end{equation}
where $w$ is a given parameter that approximates the weight of each object. Note that such weight estimate can be looked up using architecture standards such as in 
\cite{ramsey2007architectural}. Hence, if a room space $R_i$ has $n_i$ objects, then the total effort to re-arrange the space is
\begin{equation}
    E(R_i,\Theta_i) = \sum_{k=1}^{n_i}w_k\|G(O_{i,k},\theta_{i,k})\|_t,
\end{equation}
where $\Theta_i = \{\theta_{i,1},\cdots, \theta_{i,n_i} \}$ denotes the collection of $n_i$ rigid-body motion parameters.

Since solving for the optimal object transformation is an NP-Hard problem, in this paper, we will demonstrate a heuristic-based but practical algorithm to optimize it in a step-by-step greedy fashion.

\begin{equation}
\min\sum_{i=1}^{m}E(R_i,\Theta^s_i)\quad \mbox{subj. to}\quad K^s(\bigcap_{i=1}^{m} G(S_i, \theta^s_i)) \mbox{ increases 10\%},
\end{equation}
where $K^s$ indicates the area value at the $s$-th step with respect to transformation coefficients $\Theta^s_i$ and $\theta^s_i$. The iteration would stop if the optimization cannot further increase the area of the mutual space.

\begin{figure*}
\centering
  \includegraphics[width=2\columnwidth]{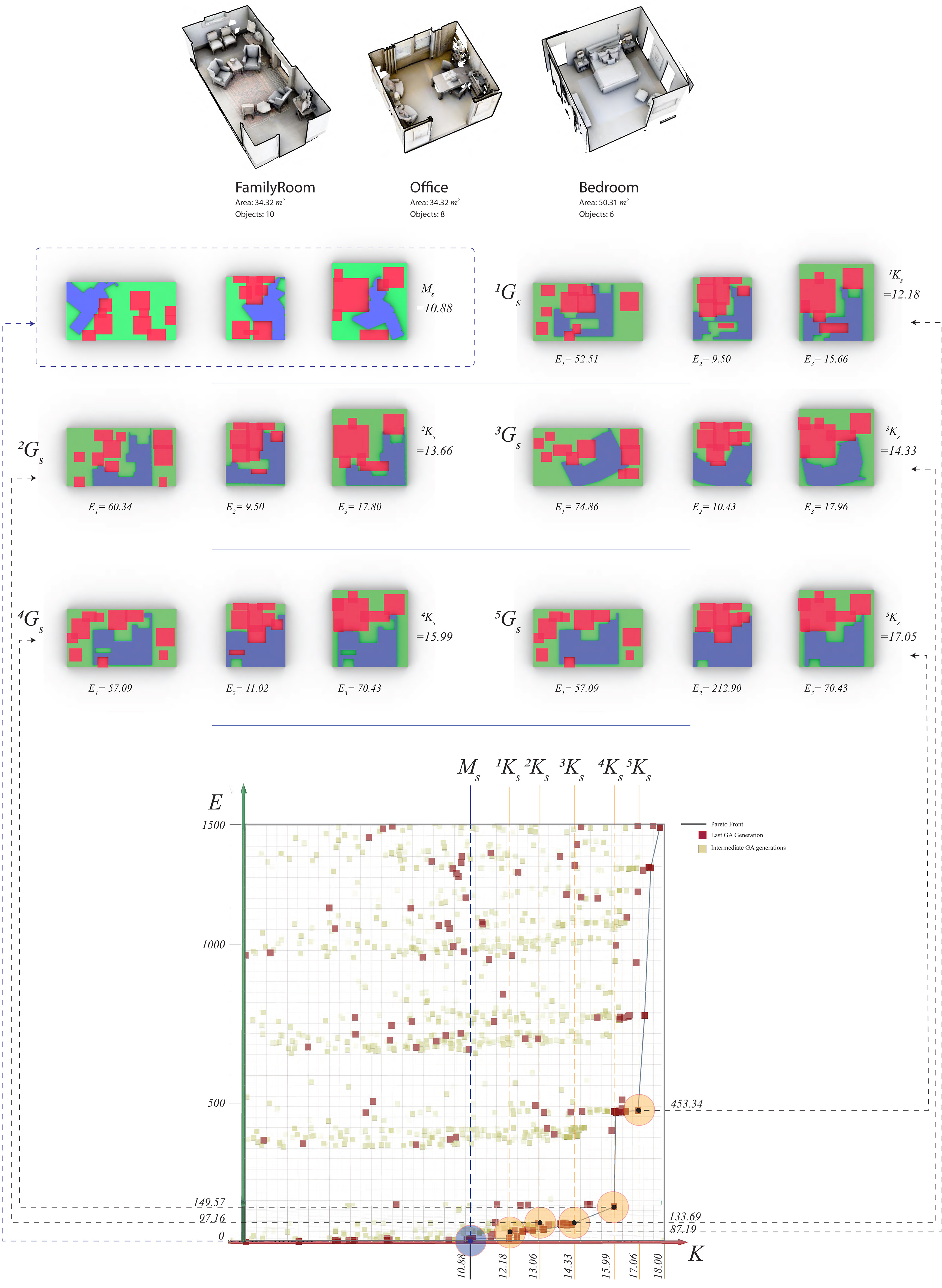}
  \caption{Furniture optimization and manipulation. In each step, a $10\%$ increase of mutual space area ($K$) is determined, while minimizing the overall effort needed ($E$) for the required transformation ($G$). }~\label{fig:FurnMo}
\end{figure*}

\section{Implementation on a 3D Scanned Dataset}

To comprehensively observe how the search and recommendation system performs given various rooms types with different spatial organizations, we take advantage of available 3D datasets to be able to experiment with large quantities of real-world case studies. We use the Matterport 3D {\cite{Chang2018}} dataset and randomly sample subsets of varying sizes of 3D scanned scenes, and perform the search and recommendation practice on each subset to observe how the mutual spaces are identified and maximized with our algorithm. Matterport 3D is a large-scale RGB-D dataset containing 90 building-scale scenes. The dataset consists of various building types with diverse architecture styles, each including numerous spatial functionalities and furniture layouts. Annotations of building elements and furniture are provided with surface reconstructions as well as 2D and 3D semantic segmentation. For our experiments, we initially exclude spaces that are not generally used for multi-user interaction (bathroom, small corridors, stairs, closet, etc.). Furthermore, we randomly group the available rooms in groups of 2, 3, and 4. We utilize the object category labels provided in the dataset as the ground truth for our semantic labeling purposes.

We implement our framework using the Rhinoceros3D (R3D) software and its development libraries. For each room, we convert the labeling data structure provided by the dataset to our proposed topological scene graph. This provides the system with bounding boxes for each object and the topological constraints for their potential rearrangement. Using such a structure, we are able to extract the standable and sittable spaces for each room based on our proposed methodology.  Figure \ref{fig:standSitMat} illustrates the available standable and sittable boundaries for two sample rooms processed by our system. We define a constant $\varepsilon_{O_{i,k}} = 70 \mbox{ cm}$ for all sittable objects. 

Next, we integrate our algorithm with a robust Strength Pareto Evolutionary Algorithm 2 (SPEA 2) \cite{zitzler2001spea2} available through the Octopus multi-objective optimization tool in R3D. The fitness function (\ref{eq:optimal-rigid-body-motion}) is used to maximize the mutual space for calculated standable spaces. Our genotype is comprised of the transformation parameters $G(F, \theta)$ of each room, allowing free movement and orientation to achieve maximum spatial consensus. Therefore, a total of $3(n-1)$ genes are allocated for the search process. This process would result in the shape, position and orientation of the maximum mutual boundary of the assigned rooms. We use a population size of 100, mutation probability of $10\%$, mutation rate of $50\%$ and crossover rate of $80\%$ for our search. As our solution integrates a genetic search, we expect the result to gradually converge to the global optimum. Figure \ref{fig:FurnMo} shows how the mutual space boundary is progressively expanded with increase of the generations in our search.

Expanding further, we extend our search by manipulating the scene with alternative furniture arrangements. As the objective goal is to achieve an increased mutual spatial boundary area with minimum effort, we calculate the $E$ based on the transformation parameters assigned to each object present in the room. However, in our current implementation, the genetic algorithm integrated in our solution is not capable of adapting dynamic genotype values, and therefore cannot update the topological values of each object ($\delta X_{max}$,  $\delta X_{min}$,  $\delta Y_{max}$, $\delta Y_{min}$) during the search process. Hence, to avoid transformations which result in physical conflicts of manipulated furniture, we penalize phenotypes that contain intersecting furniture within the scene. This penalty is added to the $E$ value, lowering the probability of such phenotypes to be selected or survive throughout the genetic generations.

\begin{figure*}
\centering
  \includegraphics[width=2\columnwidth]{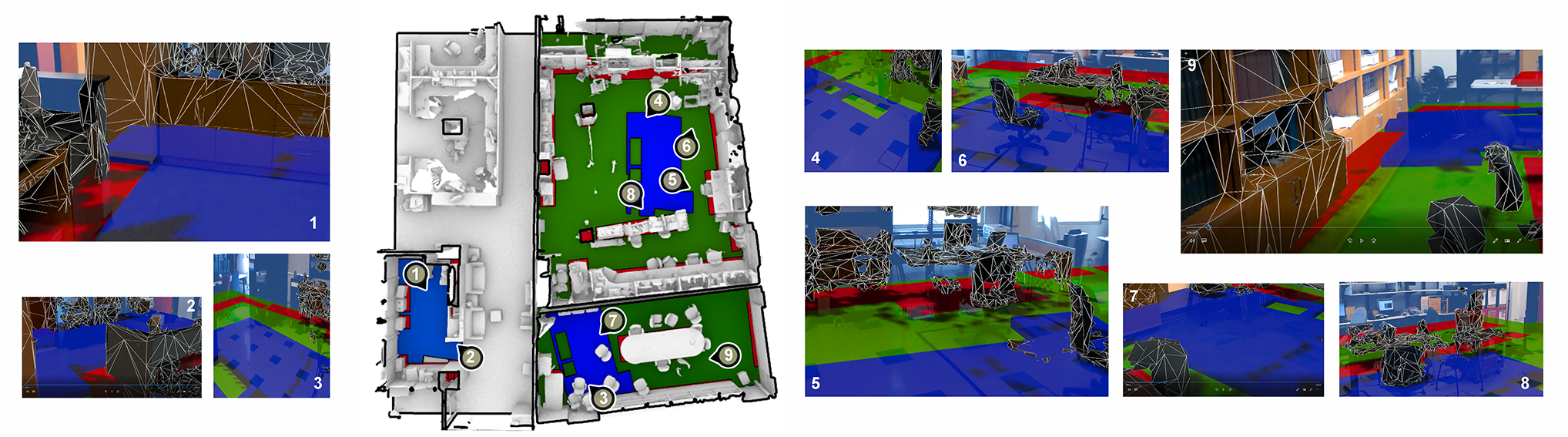}
  \caption{Screenshots from HoloLens illustrating the identified mutual boundaries as augmented overlays for three rooms: A) kitchen; B) conference room; C) robotic laboratory. Blue color indicates mutual boundaries, green color indicates standable spaces and red color indicates non-standable spaces.}~\label{fig:HoloLens}
\end{figure*}

The optimization can either be (i) triggered in separate attempts for each step ($s$), where the mutual area value ($K$) is constrained based on the resulting step value, or (ii) executed in a single attempt where minimizing $E$ and maximizing $K$ are both set as objective functions. In the latter, $M_S$ is defined as the solution which holds the largest $K$ while $E = 0$. Executing the optimization in a one-time event is also likely to require additional computational cost due to the added complexity to the solution space.

\section{Results}

Figure \ref{fig:FurnMo} illustrates our results for a furniture manipulation optimization task applied to three example rooms. A total of 34 objects are located in the rooms. To shorten our gene length we do not apply rotation transformations to objects.  We use a population size of 250, mutation probability of $10\%$, mutation rate of $50\%$ and crossover rate of $80\%$ for the scene manipulation search. We visualize the standable, sittable and mutual boundaries for each spatial expansion step. Moreover we report the corresponding $E$ for each room in the alternative furniture layout. Our results in this example indicate the solution can identify solutions which increase the maximum mutual boundary area up to 65\% more than its initial state before furniture movement.

The optimization process was able to generate a well-defined Pareto front, as seen on the bottom of Figure \ref{fig:FurnMo}, locating both the two extreme points and numerous intermediate trade-off points representing non-dominated solutions. The bottom region of the curve is flat, indicating that for a similar amount of effort, a significant increase in mutual standable area can be achieved. The trade-off frontier thus starts at point $M_S$, becoming very densely populated in its initial soft slope. This shows that for each modest increase in physical effort (that is, in moving furniture) there can be extensive gains in mutual shareable area, which is an interesting result. After $s=4$, the Pareto front becomes increasingly steep, signaling that the user would now have to significantly increase physical effort levels for modest gains in shareable area. Point ${}^{4}G_s$ thus seems to indicate a breaking point of diminishing returns.

Similar to the $M_S$ search, in smaller furniture optimization steps, the algorithm seeks solutions which are highly dependent on the transformation parameters $G(F, \theta)$ of the room itself, whereas in larger steps, we observe the algorithm correctly moving the objects to the more populated side of the room in order to increase the empty spaces in available. In rooms where objects are facing the center, and empty areas are initially located in the middle portion of the space, we see the objects being pushed towards the corners or outer perimeter of the room in order increase the initial unoccupied areas.

Due to the smaller gene size, calculating the optimal $M_S$ (maximum mutual space without furniture manipulation) executes much faster compared to $E(R_i,\Theta^s_i)$ optimization, where the complexity of the search mechanism radically increases due to the additional object transformation parameters. The speed of the $E(R_i,\Theta^s_i)$ optimization is also highly dependent on the transformation range of each object, meaning that objects in larger rooms have more movement options to choose from than those in small, constrained rooms. We observe an example of this effect in the later augmented reality experiment (Section \ref{arEx}), where the smaller space (kitchen) dominates the search process, causing the final mutual outcome between the rooms to maintain a very similar shape to the open boundaries of the smaller space. While such an effect would still provide a well-constrained problem for medium-sized rooms with multiple objects (such as the conference room), there are many possible ways of fitting the smaller space in larger rooms with open spaces (such as the robotics laboratory), resulting in an under-constrained optimization problem.

\section{Augmented Reality Visualization}\label{arEx}

To explore the usability aspect of our solution in real-world scenarios, we deploy the resulting spatial segmentation in augmented reality using the Microsoft Hololens, a mixed reality HMD. In this experiment, three types of rooms were defined as potential tele-communication spaces: (i) a conventional meeting room, where a large conference table is placed in the middle of the room and unused spaces are located around the table (ii) a robotics laboratory, where working desks and equipment are mainly located around the perimeter of the room, while some larger equipment and a few tables are disorderly positioned around the central section of the lab (iii) a kitchen space, where surrounding appliances and cabinets are present in the scene.

After the initial scan of the surrounding environment by the user of each room, the geometrical mesh data is sent to a central server for processing. This process happens in an offline manner, as the current Hololens hardware is incapable of processing the computations that our solution would require. In addition, we scan the space using a Matterport camera, and perform the semantic segmentation step using Matterport classifications to locate the bounding boxes of all the furniture located in the room. We then feed the bounding box data to our algorithm for mutual boundary search. The implementation outputs spatial coordinates for standable and sittable areas which are automatically updated in the Unity Game Engine to be rendered in the Hololenses.

Figure \ref{fig:HoloLens} shows how the spatial boundary properties are visualized within the Hololens AR experience. The red spaces indicate non standable objects, the green spaces indicate standable boundaries, and the blue spaces indicate mutual boundaries that are accessible between all users. The visualized boundaries are positioned slightly above the floor level, allowing users to identify the mutual accessible ground between their local surrounding and the remote participant's spatial constraints.

Visualizing the mutual ground within the space itself using HoloLens allows us to understand how complex the problem can be when executed in a manual fashion. Some corner spaces that are not typically used as default social areas of an certain room, may become the only required common ground for interaction with other rooms. Overcoming this spatial bias is easily executed within the algorithm; meanwhile, this may not happen so easily and instantly when individuals are left to deal with it on their own.

However, due to the limited field of view of the HoloLens, detecting non-physical boundaries placed at a lower visual height becomes difficult to follow. This issue proved  more challenging when walking closer to the non-orthogonal edges of mutual bounding area, where an individual could easily step outside the designated area. The shareable area also included a number of voids, which resulted on an inconsistent walking path inside the standable spaces. Moreover, the accuracy of the real-time mesh reconstruction in HoloLens played a critical role in calculating the required rendering occlusions for the visualized boundaries. This was mainly because the position of the the visualization was reflected close to the floor with many object placed over it, therefore failing to detect occluding objects, a fact that often misled the user in identifying whether the space was mutually accessible or not. 

\section{Conclusions}

We introduce a novel optimization and manipulation framework to generate an optimal common virtual space for interactions that mostly involve standing and sitting. Our framework further recommends movement of surrounding furniture objects that can expand the size of the mutual space with minimal physical effort. We integrated our framework with a Strength Pareto Evolutionary Algorithm for an efficient search and optimization process. The multicriteria optimization process was able to generate a well-defined Pareto front of trade-offs between maximizing mutual space and minimizing physical effort. The Pareto front is more densely populated in some sections of the frontier than others, clearly identifying the best trade-offs region and the on-start of diminishing returns. 

Furthermore, we experimented how the output solutions can be visualized using a HoloLens application. Results show that the proposed framework can effectively discover optimal shareable space for multi-user virtual interaction and thus provides better user experience compared to manually labeling shareable space, which would be a labor-intensive and imprecise workflow.  In such context, if all participants stand within the calculated mutual spatial boundaries, the line of sight between all participants will be deterministic. In addition, no remote participant will be positioned in a conflicting location for any local user and would comply to the spatial constraints for all other participants.

There are, of course, limitations to the work. First, furniture with fixed positions are not automatically detected in our current implementation. We believe such feature can be integrated with further improvements in semantic segmentation methodologies, or can be optionally specified by the user whether an object is fixed or not. In addition, the furniture weight is calculated based on standard assumptions. We envision that with the growth of spatial computing procedures, such meta-data of the surrounding environment will be customizable by the user itself and can be loaded upon each  mutual spatial search execution. Future work can comprise of integrating robust floorplanning representations with the current search mechanism to minimize computation cost and complexity. Lastly, usability studies can be conducted on how to improve the visualization strategies so participants can experience the required tele-communication functionalities while preserving the mutual spatial ground.

\acknowledgments{
We acknowledge the generous support from the following research grants: FHL Vive Center for Enhanced Reality Seed Grant, a Siemens Berkeley Industrial Partnership Grant, ONR N00014-19-1-2066.}



\begin{thebibliography}{10}

\bibitem{Abdinnour-Helm_Hadley_2000}
S.~Abdinnour-Helm and S.~W. Hadley.
\newblock Tabu search based heuristics for multi-floor facility layout.
\newblock {\em International Journal of Production Research}, 38(2):365–383,
  2000.

\bibitem{Armeni2016}
I.~Armeni, O.~Sener, A.~R. Zamir, H.~Jiang, I.~Brilakis, M.~Fischer, and
  S.~Savarese.
\newblock {3D Semantic Parsing of Large-Scale Indoor Spaces}.
\newblock {\em 2016 IEEE Conference on Computer Vision and Pattern Recognition
  (CVPR)}, pp. 1534--1543, 2016. doi: {{%
10\hspace{.1pt}\discretionary{.}{%
}{.}\hspace{.4pt}1109\discretionary{/}{%
}{/}CVPR\hspace{.1pt}\discretionary{.}{%
}{.}\hspace{.4pt}2016\hspace{.1pt}\discretionary{.}{%
}{.}\hspace{.4pt}170}}


\bibitem{balogh2010real}
T.~Balogh and P.~T. Kov{\'a}cs.
\newblock Real-time 3d light field transmission.
\newblock In {\em Real-Time Image and Video Processing 2010}, vol. 7724, p.
  772406. International Society for Optics and Photonics, 2010.

\bibitem{bardram2005activity}
E.~Bardram.
\newblock Activity-based computing: support for mobility and collaboration in
  ubiquitous computing.
\newblock {\em Personal and Ubiquitous Computing}, 9(5):312--322, 2005.

\bibitem{Beck2013}
S.~Beck, A.~Kunert, A.~Kulik, and B.~Froehlich.
\newblock {Immersive group-to-group telepresence}.
\newblock {\em IEEE Transactions on Visualization and Computer Graphics},
  19(4):616--625, 2013. doi: {{%
10\hspace{.1pt}\discretionary{.}{%
}{.}\hspace{.4pt}1109\discretionary{/}{%
}{/}TVCG\hspace{.1pt}\discretionary{.}{%
}{.}\hspace{.4pt}2013\hspace{.1pt}\discretionary{.}{%
}{.}\hspace{.4pt}33}}


\bibitem{beck2013immersive}
S.~Beck, A.~Kunert, A.~Kulik, and B.~Froehlich.
\newblock Immersive group-to-group telepresence.
\newblock {\em IEEE Transactions on Visualization and Computer Graphics},
  19(4):616--625, 2013.

\bibitem{benko2012miragetable}
H.~Benko, R.~Jota, and A.~Wilson.
\newblock Miragetable: freehand interaction on a projected augmented reality
  tabletop.
\newblock In {\em Proceedings of the SIGCHI conference on human factors in
  computing systems}, pp. 199--208. ACM, 2012.

\bibitem{blanche2010holographic}
P.-A. Blanche, A.~Bablumian, R.~Voorakaranam, C.~Christenson, W.~Lin, T.~Gu,
  D.~Flores, P.~Wang, W.-Y. Hsieh, M.~Kathaperumal, et~al.
\newblock Holographic three-dimensional telepresence using large-area
  photorefractive polymer.
\newblock {\em Nature}, 468(7320):80, 2010.

\bibitem{bolte2015subliminal}
B.~Bolte and M.~Lappe.
\newblock Subliminal reorientation and repositioning in immersive virtual
  environments using saccadic suppression.
\newblock {\em IEEE transactions on visualization and computer graphics},
  21(4):545--552, 2015.

\bibitem{bruder2015cognitive}
G.~Bruder, P.~Lubos, and F.~Steinicke.
\newblock Cognitive resource demands of redirected walking.
\newblock {\em IEEE transactions on visualization and computer graphics},
  21(4):539--544, 2015.

\bibitem{bruder2011tuning}
G.~Bruder, F.~Steinicke, P.~Wieland, and M.~Lappe.
\newblock Tuning self-motion perception in virtual reality with visual
  illusions.
\newblock {\em IEEE Transactions on Visualization and Computer Graphics},
  18(7):1068--1078, 2011.

\bibitem{caldas2019design}
L.~Caldas and M.~Keshavarzi.
\newblock Design immersion and virtual presence.
\newblock {\em Technology| Architecture+ Design}, 3(2):249--251, 2019.

\bibitem{Chang2018}
A.~Chang, A.~Dai, T.~Funkhouser, M.~Halber, M.~Niebner, M.~Savva, S.~Song,
  A.~Zeng, and Y.~Zhang.
\newblock {Matterport3D: Learning from RGB-D data in indoor environments}.
\newblock In {\em Proceedings - 2017 International Conference on 3D Vision, 3DV
  2017}, pp. 667--676, 2018. doi: {{%
10\hspace{.1pt}\discretionary{.}{%
}{.}\hspace{.4pt}1109\discretionary{/}{%
}{/}3DV\hspace{.1pt}\discretionary{.}{%
}{.}\hspace{.4pt}2017\hspace{.1pt}\discretionary{.}{%
}{.}\hspace{.4pt}00081}}


\bibitem{Chang_Chang_Wu_Wu_2000}
Y.-C. Chang, Y.-W. Chang, G.-M. Wu, and S.-W. Wu.
\newblock B*-trees: a new representation for non-slicing floorplans.
\newblock In {\em Proceedings 37th Design Automation Conference}, p. 458–463,
  2000. doi: {{%
10\hspace{.1pt}\discretionary{.}{%
}{.}\hspace{.4pt}1109\discretionary{/}{%
}{/}DAC\hspace{.1pt}\discretionary{.}{%
}{.}\hspace{.4pt}2000\hspace{.1pt}\discretionary{.}{%
}{.}\hspace{.4pt}855354}}


\bibitem{chen2008vlsi}
G.~Chen, W.~Guo, H.~Cheng, X.~Fen, and X.~Fang.
\newblock Vlsi floorplanning based on particle swarm optimization.
\newblock In {\em 2008 3rd International Conference on Intelligent System and
  Knowledge Engineering}, vol.~1, pp. 1020--1025. IEEE, 2008.

\bibitem{Cheng_1992}
M.~Y. Cheng.
\newblock Automated site layout of temporary construction facilities
  using-enhanced geographic information systems (gis).
\newblock {\em Ph. D. Disst., Depart. of Civil Engineering, University of Texas
  at Austin, Texas, USA}, 1992.

\bibitem{clark1991grounding}
H.~H. Clark, S.~E. Brennan, et~al.
\newblock Grounding in communication.
\newblock {\em Perspectives on socially shared cognition}, 13(1991):127--149,
  1991.

\bibitem{eisenmann1998generic}
H.~Eisenmann, F.~M. Johannes, and F.~M. Johannes.
\newblock Generic global placement and floorplanning.
\newblock In {\em Proceedings of the 35th annual Design Automation Conference},
  pp. 269--274. ACM, 1998.

\bibitem{Fuchs2014}
H.~Fuchs, A.~State, and J.~C. Bazin.
\newblock {Immersive 3D telepresence}.
\newblock {\em Computer}, 47(7):46--52, 2014. doi: {{%
10\hspace{.1pt}\discretionary{.}{%
}{.}\hspace{.4pt}1109\discretionary{/}{%
}{/}MC\hspace{.1pt}\discretionary{.}{%
}{.}\hspace{.4pt}2014\hspace{.1pt}\discretionary{.}{%
}{.}\hspace{.4pt}185}}


\bibitem{gross2003blue}
M.~Gross, S.~W{\"u}rmlin, M.~Naef, E.~Lamboray, C.~Spagno, A.~Kunz,
  E.~Koller-Meier, T.~Svoboda, L.~Van~Gool, et~al.
\newblock blue-c: a spatially immersive display and 3d video portal for
  telepresence.
\newblock In {\em ACM Transactions on Graphics (TOG)}, vol.~22, pp. 819--827.
  ACM, 2003.

\bibitem{Guo_Cheng_Yoshimura_2003}
P.-N. Guo, C.-K. Cheng, and T.~Yoshimura.
\newblock An o-tree representation of non-slicing floorplan and its
  applications.
\newblock In {\em Proceedings 1999 Design Automation Conference (Cat. No.
  99CH36361)}, p. 268–273. IEEE, 2003. doi: {{%
10\hspace{.1pt}\discretionary{.}{%
}{.}\hspace{.4pt}1145\discretionary{/}{%
}{/}309847\hspace{.1pt}\discretionary{.}{%
}{.}\hspace{.4pt}309928}}


\bibitem{gwee1999ga}
B.-H. Gwee and M.-H. Lim.
\newblock A ga with heuristic-based decoder for ic floorplanning.
\newblock {\em INTEGRATION, the VLSI journal}, 28(2):157--172, 1999.

\bibitem{interrante2007seven}
V.~Interrante, B.~Ries, and L.~Anderson.
\newblock Seven league boots: A new metaphor for augmented locomotion through
  moderately large scale immersive virtual environments.
\newblock In {\em 2007 IEEE Symposium on 3D User interfaces}. IEEE, 2007.

\bibitem{Jo_Gero_1998}
J.~H. Jo and J.~S. Gero.
\newblock Space layout planning using an evolutionary approach.
\newblock {\em Artificial Intelligence in Engineering}, 12(3):149–162, Jul
  1998. doi: {{%
10\hspace{.1pt}\discretionary{.}{%
}{.}\hspace{.4pt}1016\discretionary{/}{%
}{/}S0954\discretionary{%
}{-}{-}1810\discretionary{%
}{(}{(}97\discretionary{)}{%
}{)}00037\discretionary{%
}{-}{-}X}}


\bibitem{jones2009achieving}
A.~Jones, M.~Lang, G.~Fyffe, X.~Yu, J.~Busch, I.~McDowall, M.~Bolas, and
  P.~Debevec.
\newblock Achieving eye contact in a one-to-many 3d video teleconferencing
  system.
\newblock In {\em ACM Transactions on Graphics (TOG)}, vol.~28, p.~64. ACM,
  2009.

\bibitem{jones2014roomalive}
B.~Jones, R.~Sodhi, M.~Murdock, R.~Mehra, H.~Benko, A.~Wilson, E.~Ofek,
  B.~MacIntyre, N.~Raghuvanshi, and L.~Shapira.
\newblock Roomalive: magical experiences enabled by scalable, adaptive
  projector-camera units.
\newblock In {\em Proceedings of the 27th annual ACM symposium on User
  interface software and technology}, pp. 637--644. ACM, 2014.

\bibitem{kahng2011vlsi}
A.~B. Kahng, J.~Lienig, I.~L. Markov, and J.~Hu.
\newblock {\em VLSI physical design: from graph partitioning to timing
  closure}.
\newblock Springer Science \& Business Media, 2011.

\bibitem{kaur2014enhanced}
P.~Kaur.
\newblock An enhanced algorithm for floorplan design using hybrid ant colony
  and particle swarm optimization.
\newblock {\em Int. J. Res. Appl. Sci. Eng. Technol}, 2:473--477, 2014.

\bibitem{keshavarzi2019affordance}
M.~Keshavarzi, M.~Wu, M.~N. Chin, R.~N. Chin, and A.~Y. Yang.
\newblock Affordance analysis of virtual and augmented reality mediated
  communication.
\newblock {\em arXiv preprint arXiv:1904.04723}, 2019.

\bibitem{kim2012telehuman}
K.~Kim, J.~Bolton, A.~Girouard, J.~Cooperstock, and R.~Vertegaal.
\newblock Telehuman: effects of 3d perspective on gaze and pose estimation with
  a life-size cylindrical telepresence pod.
\newblock In {\em Proceedings of the SIGCHI Conference on Human Factors in
  Computing Systems}, pp. 2531--2540. ACM, 2012.

\bibitem{kirkpatrick1983optimization}
S.~Kirkpatrick, C.~D. Gelatt, and M.~P. Vecchi.
\newblock Optimization by simulated annealing.
\newblock {\em science}, 220(4598):671--680, 1983.

\bibitem{kiyota2005simulated}
K.~Kiyota and K.~Fujiyoshi.
\newblock Simulated annealing search through general structure floor plans
  using sequence-pair.
\newblock {\em Electronics and Communications in Japan (Part III: Fundamental
  Electronic Science)}, 88(6):28--38, 2005.

\bibitem{kurillo2008immersive}
G.~Kurillo, R.~Bajcsy, K.~Nahrsted, and O.~Kreylos.
\newblock Immersive 3d environment for remote collaboration and training of
  physical activities.
\newblock In {\em 2008 IEEE Virtual Reality Conference}, pp. 269--270. IEEE,
  2008.

\bibitem{Kuster2012}
C.~Kuster, N.~Ranieri, Agustina, H.~Zimmer, J.~C. Bazin, C.~Sun, T.~Popa, and
  M.~Gross.
\newblock {Towards next generation 3D teleconferencing systems}.
\newblock In {\em 3DTV-Conference}, number Section 2, pp. 1--4. IEEE, oct 2012.
  doi: {{%
10\hspace{.1pt}\discretionary{.}{%
}{.}\hspace{.4pt}1109\discretionary{/}{%
}{/}3DTV\hspace{.1pt}\discretionary{.}{%
}{.}\hspace{.4pt}2012\hspace{.1pt}\discretionary{.}{%
}{.}\hspace{.4pt}6365454}}


\bibitem{lang2019virtual}
V.~Lang, W.~Liang, and L.-F. Yu.
\newblock Virtual agent positioning driven by scene semantics in mixed reality.
\newblock In {\em 2019 IEEE Conference on Virtual Reality and 3D User
  Interfaces (VR)}, pp. 767--775. IEEE, 2019.

\bibitem{LEE200121}
B.~P.~H. Lee.
\newblock {Mutual knowledge, background knowledge and shared beliefs: Their
  roles in establishing common ground}.
\newblock {\em Journal of Pragmatics}, 33(1):21--44, 2001. doi: {{%
10\hspace{.1pt}\discretionary{.}{%
}{.}\hspace{.4pt}1016\discretionary{/}{%
}{/}S0378\discretionary{%
}{-}{-}2166\discretionary{%
}{(}{(}99\discretionary{)}{%
}{)}00128\discretionary{%
}{-}{-}9}}


\bibitem{Lehment2014}
N.~H. Lehment, D.~Merget, and G.~Rigoll.
\newblock {Creating automatically aligned consensus realities for AR
  videoconferencing}.
\newblock {\em ISMAR 2014 - IEEE International Symposium on Mixed and Augmented
  Reality - Science and Technology 2014, Proceedings}, (September):201--206,
  2014.

\bibitem{lin2002efficient}
C.-T. Lin, D.-S. Chen, and Y.-W. Wang.
\newblock An efficient genetic algorithm for slicing floorplan area
  optimization.
\newblock In {\em 2002 IEEE International Symposium on Circuits and Systems.
  Proceedings (Cat. No. 02CH37353)}, vol.~2, pp. II--II. IEEE, 2002.

\bibitem{lin2005tcg}
J.-M. Lin and Y.-W. Chang.
\newblock Tcg: A transitive closure graph-based representation for general
  floorplans.
\newblock {\em IEEE transactions on very large scale integration (VLSI)
  systems}, 13(2):288--292, 2005.

\bibitem{Liu2018}
C.~Liu, J.~Wu, and Y.~Furukawa.
\newblock {FloorNet: A Unified Framework for Floorplan Reconstruction from 3D
  Scans}.
\newblock pp. 1--18, 2018.

\bibitem{Lombardi}
S.~Lombardi, J.~Saragih, T.~Simon, and Y.~Sheikh.
\newblock Deep appearance models for face rendering.
\newblock {\em ACM Trans. Graph.}, 37(4):68:1--68:13, July 2018. doi: {{%
10\hspace{.1pt}\discretionary{.}{%
}{.}\hspace{.4pt}1145\discretionary{/}{%
}{/}3197517\hspace{.1pt}\discretionary{.}{%
}{.}\hspace{.4pt}3201401}}


\bibitem{luff1998mobility}
P.~Luff and C.~Heath.
\newblock Mobility in collaboration.
\newblock In {\em CSCW}, vol.~98, pp. 305--314, 1998.

\bibitem{ma2001vlsi}
Y.~Ma, S.~Dong, X.~Hong, Y.~Cai, C.-K. Cheng, and J.~Gu.
\newblock Vlsi floorplanning with boundary constraints based on corner block
  list.
\newblock In {\em Proceedings of the 2001 Asia and South Pacific Design
  Automation Conference}, pp. 509--514. ACM, 2001.

\bibitem{maimone2011encumbrance}
A.~Maimone and H.~Fuchs.
\newblock Encumbrance-free telepresence system with real-time 3d capture and
  display using commodity depth cameras.
\newblock In {\em 2011 10th IEEE International Symposium on Mixed and Augmented
  Reality}, pp. 137--146. IEEE, 2011.

\bibitem{maimone2012real}
A.~Maimone and H.~Fuchs.
\newblock Real-time volumetric 3d capture of room-sized scenes for
  telepresence.
\newblock In {\em 2012 3DTV-Conference: The True Vision-Capture, Transmission
  and Display of 3D Video (3DTV-CON)}, pp. 1--4. IEEE, 2012.

\bibitem{maimone2013general}
A.~Maimone, X.~Yang, N.~Dierk, A.~State, M.~Dou, and H.~Fuchs.
\newblock General-purpose telepresence with head-worn optical see-through
  displays and projector-based lighting.
\newblock In {\em 2013 IEEE Virtual Reality (VR)}, pp. 23--26. IEEE, 2013.

\bibitem{matusik20043d}
W.~Matusik and H.~Pfister.
\newblock 3d tv: a scalable system for real-time acquisition, transmission, and
  autostereoscopic display of dynamic scenes.
\newblock In {\em ACM Transactions on Graphics (TOG)}, vol.~23, pp. 814--824.
  ACM, 2004.

\bibitem{moni2009vlsi}
D.~J. Moni and S.~Arumugam.
\newblock Vlsi floorplanning based on hybrid particle swarm optimization.
\newblock {\em Karunya Journal of Research}, 1(1):111--121, 2009.

\bibitem{nagano2013autostereoscopic}
K.~Nagano, A.~Jones, J.~Liu, J.~Busch, X.~Yu, M.~Bolas, and P.~Debevec.
\newblock An autostereoscopic projector array optimized for 3d facial display.
\newblock In {\em ACM SIGGRAPH 2013 Emerging Technologies}, p.~3. ACM, 2013.

\bibitem{nakatake1997module}
S.~Nakatake, K.~Fujiyoshi, H.~Murata, and Y.~Kajitani.
\newblock Module placement on bsg-structure and ic layout applications.
\newblock In {\em Proceedings of the 1996 IEEE/ACM international conference on
  Computer-aided design}, pp. 484--491. IEEE Computer Society, 1997.

\bibitem{nakaya2000adaptive}
S.~Nakaya, T.~Koide, and S.~Wakabayashi.
\newblock An adaptive genetic algorithm for vlsi floorplanning based on
  sequence-pair.
\newblock In {\em 2000 IEEE International Symposium on Circuits and Systems.
  Emerging Technologies for the 21st Century. Proceedings (IEEE Cat No.
  00CH36353)}, vol.~3, pp. 65--68. IEEE, 2000.

\bibitem{narang2018simulating}
S.~Narang, A.~Best, and D.~Manocha.
\newblock Simulating movement interactions between avatars \& agents in virtual
  worlds using human motion constraints.
\newblock In {\em 2018 IEEE Conference on Virtual Reality and 3D User
  Interfaces (VR)}, pp. 9--16. IEEE, 2018.

\bibitem{Nilsson2018}
N.~C. Nilsson, T.~Peck, G.~Bruder, E.~Hodgson, S.~Serafin, M.~Whitton,
  F.~Steinicke, and E.~S. Rosenberg.
\newblock {15 Years of Research on Redirected Walking in Immersive Virtual
  Environments}.
\newblock {\em IEEE Computer Graphics and Applications}, 38(2):44--56, 2018.
  doi: {{%
10\hspace{.1pt}\discretionary{.}{%
}{.}\hspace{.4pt}1109\discretionary{/}{%
}{/}MCG\hspace{.1pt}\discretionary{.}{%
}{.}\hspace{.4pt}2018\hspace{.1pt}\discretionary{.}{%
}{.}\hspace{.4pt}111125628}}


\bibitem{Orts-Escolano2017}
S.~Orts-Escolano, M.~Dou, V.~Tankovich, C.~Loop, Q.~Cai, P.~A. Chou,
  S.~Mennicken, J.~Valentin, V.~Pradeep, S.~Wang, S.~B. Kang, C.~Rhemann,
  P.~Kohli, Y.~Lutchyn, C.~Keskin, S.~Izadi, S.~Fanello, W.~Chang, A.~Kowdle,
  Y.~Degtyarev, D.~Kim, P.~L. Davidson, and S.~Khamis.
\newblock {Holoportation}.
\newblock pp. 741--754, 2017. doi: {{%
10\hspace{.1pt}\discretionary{.}{%
}{.}\hspace{.4pt}1145\discretionary{/}{%
}{/}2984511\hspace{.1pt}\discretionary{.}{%
}{.}\hspace{.4pt}2984517}}


\bibitem{Osman_Georgy_Ibrahim_2003}
H.~M. Osman, M.~E. Georgy, and M.~E. Ibrahim.
\newblock A hybrid cad-based construction site layout planning system using
  genetic algorithms.
\newblock {\em Automation in Construction}, 12(6):749–764, 2003. doi: {{%
10\hspace{.1pt}\discretionary{.}{%
}{.}\hspace{.4pt}1016\discretionary{/}{%
}{/}S0926\discretionary{%
}{-}{-}5805\discretionary{%
}{(}{(}03\discretionary{)}{%
}{)}00058\discretionary{%
}{-}{-}X}}


\bibitem{peck2011evaluation}
T.~C. Peck, H.~Fuchs, and M.~C. Whitton.
\newblock An evaluation of navigational ability comparing redirected free
  exploration with distractors to walking-in-place and joystick locomotio
  interfaces.
\newblock In {\em 2011 IEEE Virtual Reality Conference}, pp. 55--62. IEEE,
  2011.

\bibitem{Pejsa:2016:REL:2818048.2819965}
T.~Pejsa, J.~Kantor, H.~Benko, E.~Ofek, and A.~Wilson.
\newblock Room2room: Enabling life-size telepresence in a projected augmented
  reality environment.
\newblock In {\em Proceedings of the 19th ACM Conference on Computer-Supported
  Cooperative Work \& Social Computing}, CSCW '16, pp. 1716--1725. ACM, New
  York, NY, USA, 2016. doi: {{%
10\hspace{.1pt}\discretionary{.}{%
}{.}\hspace{.4pt}1145\discretionary{/}{%
}{/}2818048\hspace{.1pt}\discretionary{.}{%
}{.}\hspace{.4pt}2819965}}


\bibitem{Qi2016}
C.~R. Qi, H.~Su, M.~Niessner, A.~Dai, M.~Yan, and L.~J. Guibas.
\newblock {Volumetric and Multi-View CNNs for Object Classification on 3D
  Data}.
\newblock 2016. doi: {{%
10\hspace{.1pt}\discretionary{.}{%
}{.}\hspace{.4pt}1109\discretionary{/}{%
}{/}CVPR\hspace{.1pt}\discretionary{.}{%
}{.}\hspace{.4pt}2016\hspace{.1pt}\discretionary{.}{%
}{.}\hspace{.4pt}609}}


\bibitem{ramsey2007architectural}
C.~G. Ramsey.
\newblock {\em Architectural graphic standards}.
\newblock John Wiley \& Sons, 2007.

\bibitem{Razzaque2001}
S.~Razzaque, Z.~Kohn, and M.~C. Whitton.
\newblock {Redirected Walking}.
\newblock {\em Proceedings of EUROGRAPHICS}, pp. 289--294, 2001.

\bibitem{rebaudengo1996gallo}
M.~Rebaudengo and M.~S. Reorda.
\newblock Gallo: A genetic algorithm for floorplan area optimization.
\newblock {\em IEEE Transactions on Computer-Aided Design of Integrated
  Circuits and Systems}, 15(8):943--951, 1996.

\bibitem{sowmya2013minimization}
B.~Sowmya and M.~Sunil.
\newblock Minimization of floorplanning area and wire length interconnection
  using particle swarm optimization.
\newblock {\em International Journal of Emerging Technology and Advanced
  Engineering}, 3(8), 2013.

\bibitem{Suma2011}
E.~A. Suma, S.~Clark, D.~Krum, S.~Finkelstein, M.~Bolas, and Z.~Warte.
\newblock {Leveraging change blindness for redirection in virtual
  environments}.
\newblock In {\em Proceedings - IEEE Virtual Reality}, pp. 159--166, 2011. doi:
  {{%
10\hspace{.1pt}\discretionary{.}{%
}{.}\hspace{.4pt}1109\discretionary{/}{%
}{/}VR\hspace{.1pt}\discretionary{.}{%
}{.}\hspace{.4pt}2011\hspace{.1pt}\discretionary{.}{%
}{.}\hspace{.4pt}5759455}}


\bibitem{Suma2012}
E.~A. Suma, Z.~Lipps, S.~Finkelstein, D.~M. Krum, and M.~Bolas.
\newblock {Impossible spaces: Maximizing natural walking in virtual
  environments with self-overlapping architecture}.
\newblock {\em IEEE Transactions on Visualization and Computer Graphics},
  18(4):555--564, 2012. doi: {{%
10\hspace{.1pt}\discretionary{.}{%
}{.}\hspace{.4pt}1109\discretionary{/}{%
}{/}TVCG\hspace{.1pt}\discretionary{.}{%
}{.}\hspace{.4pt}2012\hspace{.1pt}\discretionary{.}{%
}{.}\hspace{.4pt}47}}


\bibitem{sun2006floorplanning}
T.-Y. Sun, S.-T. Hsieh, H.-M. Wang, and C.-W. Lin.
\newblock Floorplanning based on particle swarm optimization.
\newblock In {\em IEEE Computer Society Annual Symposium on Emerging VLSI
  Technologies and Architectures (ISVLSI'06)}, pp. 5--pp. IEEE, 2006.

\bibitem{tanikawa2005real}
T.~Tanikawa, Y.~Suzuki, K.~Hirota, and M.~Hirose.
\newblock Real world video avatar: real-time and real-size transmission and
  presentation of human figure.
\newblock In {\em Proceedings of the 2005 international conference on Augmented
  tele-existence}, pp. 112--118. ACM, 2005.

\bibitem{Tommelein_Levitt_Hayes-Roth_Confrey_1991}
I.~D. Tommelein, R.~E. Levitt, B.~Hayes-Roth, and T.~Confrey.
\newblock Sightplan experiments: alternate strategies for site layout design.
\newblock {\em Computing in Civil Engineering}, 5(1):42–63, 1991. doi: {{%
10\hspace{.1pt}\discretionary{.}{%
}{.}\hspace{.4pt}1007\discretionary{/}{%
}{/}BF01927759}}


\bibitem{towles20023d}
H.~Towles, W.-C. Chen, R.~Yang, S.-U. Kum, , H.~Fuchs, N.~Kelshikar,
  J.~Mulligan, K.~Daniilidis, L.~Holden, B.~Zeleznik, A.~Sadagic, and
  J.~Lanier.
\newblock 3d tele-collaboration over internet2.
\newblock In {\em In: International Workshop on Immersive Telepresence, Juan
  Les Pins}. Citeseer, 2002.

\bibitem{Vasylevska2013}
K.~Vasylevska, H.~Kaufmann, M.~Bolas, and E.~A. Suma.
\newblock {Flexible spaces: Dynamic layout generation for infinite walking in
  virtual environments}.
\newblock In {\em IEEE Symposium on 3D User Interface 2013, 3DUI 2013 -
  Proceedings}, pp. 39--42, 2013. doi: {{%
10\hspace{.1pt}\discretionary{.}{%
}{.}\hspace{.4pt}1109\discretionary{/}{%
}{/}3DUI\hspace{.1pt}\discretionary{.}{%
}{.}\hspace{.4pt}2013\hspace{.1pt}\discretionary{.}{%
}{.}\hspace{.4pt}6550194}}


\bibitem{Wang:2009:EDA:2843514}
L.-T. Wang, Y.-W. Chang, and K.-T.~T. Cheng, eds.
\newblock {\em Electronic Design Automation: Synthesis, Verification, and
  Test}.
\newblock Morgan Kaufmann Publishers Inc., San Francisco, CA, USA, 2009.

\bibitem{wei2019vr}
S.-E. Wei, J.~Saragih, T.~Simon, A.~W. Harley, S.~Lombardi, M.~Perdoch,
  A.~Hypes, D.~Wang, H.~Badino, and Y.~Sheikh.
\newblock Vr facial animation via multiview image translation.
\newblock {\em ACM Transactions on Graphics (TOG)}, 38(4):67, 2019.

\bibitem{wen2000toward}
W.-C. Wen, H.~Towles, L.~Nyland, G.~Welch, and H.~Fuchs.
\newblock Toward a compelling sensation of telepresence: Demonstrating a portal
  to a distant (static) office.
\newblock In {\em Proceedings Visualization 2000. VIS 2000 (Cat. No.
  00CH37145)}, pp. 327--333. IEEE, 2000.

\bibitem{williams2007exploring}
B.~Williams, G.~Narasimham, B.~Rump, T.~P. McNamara, T.~H. Carr, J.~Rieser, and
  B.~Bodenheimer.
\newblock Exploring large virtual environments with an hmd when physical space
  is limited.
\newblock In {\em Proceedings of the 4th symposium on Applied perception in
  graphics and visualization}, pp. 41--48, 2007.

\bibitem{Wong:1986:NAF:318013.318030}
D.~F. Wong and C.~L. Liu.
\newblock {A New Algorithm for Floorplan Design}.
\newblock In {\em Proceedings of the 23rd ACM/IEEE Design Automation
  Conference}, DAC '86, pp. 101--107. IEEE Press, Piscataway, NJ, USA, 1986.

\bibitem{xiaogang2002vlsi}
W.~Xiaogang et~al.
\newblock Vlsi floorplanning method based on genetic algorithms [j].
\newblock {\em Microprocessors}, (1):1, 2002.

\bibitem{zhang2013viewport}
C.~Zhang, Q.~Cai, P.~A. Chou, Z.~Zhang, and R.~Martin-Brualla.
\newblock Viewport: A distributed, immersive teleconferencing system with
  infrared dot pattern.
\newblock {\em IEEE MultiMedia}, 20(1):17--27, 2013.

\bibitem{zitzler2001spea2}
E.~Zitzler, M.~Laumanns, and L.~Thiele.
\newblock Spea2: Improving the strength pareto evolutionary algorithm.
\newblock {\em TIK-report}, 103, 2001.

\end{thebibliography}

\end{document}